\documentclass[conference]{IEEEtran}
\usepackage{cite}
\usepackage{amsmath,amssymb,amsfonts}
\usepackage{algorithmic}
\usepackage{graphicx}
\usepackage{textcomp}
\usepackage{xcolor}
\usepackage{tikz}
\usepackage[hidelinks]{hyperref}
\usetikzlibrary{positioning,arrows.meta}
\usepackage[table]{xcolor}
\usepackage{booktabs}
\usepackage{graphicx}
\usepackage{dblfloatfix}
\usepackage{svg}
\usepackage{pgfplots}
\usepackage{comment}
\usepackage[caption=false,font=footnotesize]{subfig}
\usepgfplotslibrary{groupplots}
\usetikzlibrary{calc}

\usetikzlibrary{arrows.meta,positioning}
\renewcommand{\IEEEauthorrefmark}[1]{\textsuperscript{#1}}

\def\BibTeX{{\rm B\kern-.05em{\sc i\kern-.025em b}\kern-.08em
T\kern-.1667em\lower.7ex\hbox{E}\kern-.125emX}}

\makeatletter

\def\blfootnote{\gdef\@thefnmark{}\@footnotetext}

\makeatother

\begin{document}

%\title{Cross-Attention Transformer for Joint Multi-Receiver Uplink Neural Decoding}

%\title{Scalable Cross-Attention Transformer for Multi-AP OFDM Uplink Reception}

\title{Scalable Cross-Attention Transformer for Cooperative Multi-AP OFDM Uplink Reception}

\author{
    \IEEEauthorblockN{Xavier Tardy\IEEEauthorrefmark{1,2}, Grégoire Lefebvre\IEEEauthorrefmark{2}, Apostolos Kountouris\IEEEauthorrefmark{2}, Haïfa Farès\IEEEauthorrefmark{1}, Amor Nafkha\IEEEauthorrefmark{1}}
    \IEEEauthorblockA{\IEEEauthorrefmark{1}IETR - UMR CNRS 6164, CentraleSupélec, avenue de la Boulaie - CS 47601 35576 CESSON-SEVIGNE Cedex, France\\
    Email: firstname.lastname@centralesupelec.fr}
    \IEEEauthorblockA{\IEEEauthorrefmark{2}Orange Research, Grenoble, France\\
    Email: firstname.lastname@orange.com}
}

\maketitle
\begin{abstract}
We propose a cross-attention Transformer for joint decoding of uplink OFDM signals received by multiple coordinated access points. A shared per-receiver encoder learns the time–frequency structure of each grid, and a token-wise cross-attention module fuses the receivers to produce log-likelihood ratios for a standard channel decoder, without explicit channel estimates. Trained with a bit-metric objective, the model adapts its fusion to per-receiver reliability and remains robust to degraded links, strong frequency selectivity, and sparse pilots. Over realistic wireless channels representative of Wi-Fi deployments, it outperforms classical pipelines and strong neural baselines, and often approaches a global MMSE receiver with perfect CSI. It remains compact and computationally efficient, making it suitable for next-generation coordinated Wi-Fi receivers.
\end{abstract}


\section{Introduction}

\IEEEPARstart{T}{he}\blfootnote{This work was supported in part by Bpifrance under the France 2030 i‑Démo program (Wi-FIP project, 2023–2026, Grant I-DEMO-52255).} ongoing evolution of wireless standards and deployments—exemplified by recent advances in IEEE 802.11be (Wi-Fi 7) and the emerging P802.11bn (Wi-Fi 8) standard \cite{noauthor_p80211bn_2024}—is driving unprecedented demands for throughput, reliability, and multi-link coordination, making cooperative uplink reception a key enabler for coverage and interference robustness in OFDM systems.
Coordinated multi‑AP reception, in the spirit of cell-free \cite{demir_foundations_2021} or Coordinated Multi-Point (CoMP) architectures \cite{gesbert_multicell_2010}, exploits geographically diverse observations of the same uplink transmission to enhance spatial diversity and mitigate local traffic hotspots. Joint processing of OFDM grids enables BER and robustness gains via coherent multi-AP combining, and recent fronthaul/edge advances lower implementation barriers.
However, conventional receiver pipelines remain a limiting factor in realizing these gains. These pipelines typically consist of three stages: pilot-based channel estimation, equalization, and demapping. Simple estimators like Least Squares (LS) \cite{van_de_beek_channel_1995} are sensitive to noise, while optimal linear schemes like Linear Minimum Mean Square Error (LMMSE) \cite{biguesh_trainingbased_2006} require accurate second-order channel statistics that are often unavailable or quickly outdated in non-stationary environments. Moreover, performing these steps independently at each AP ignores the spatial correlations that exist between APs and does not adapt the fusion process to the varying reliability of each AP. As a result, significant cooperative gains remain unexploited.

Beyond conventional linear processing, recent works have explored learned cooperative reception and equalization in cell-free and multi-AP settings, e.g., via in-context learning with sequence models \cite{zecchiN_cell-free_2024} and fully-decoupled RAN architectures with multi-point combining \cite{yu_large_2025}. However, these approaches typically do not operate on full 2D OFDM grids and often address only the equalization or hard symbol detection stage. Moreover, existing sequence-model designs based on full self-attention incur quadratic complexity in both the number of time–frequency elements in the resource grid and the number of receivers, which quickly becomes prohibitive for large OFDM blocks and dense multi-AP deployments. Recent state-space models \cite{song_-context_2025} offer linear complexity and strong 1D sequence performance, making them promising for scalable physical-layer processing. However, extending such models to jointly capture 2D time–frequency structure and cross-AP interactions is less straightforward than using cross-attention. 

Motivated by these limitations and inspired by the recent successes of machine learning for the physical layer \cite{ye_power_2018, honkala_deeprx_2021, xie_comm-transformer_2024}, we therefore adopt a Transformer-based architecture with cross-attention, which naturally handles 2D OFDM grids and multi-AP fusion and is aligned with recent large-sequence Transformer receivers for cooperative MIMO equalization in fully-decoupled RANs \cite{yu_large_2025}. Cross-attention mechanisms capture inter-AP and inter-subcarrier dependencies. Attention assigns data-dependent weights so that each time–frequency position forms a soft, content-aware combination of the most relevant neighbors (with pilots acting as anchors), while token-wise cross-attention applies the same principle across APs to achieve a fusion of multi-views in the embedding space. This approach enables fusion that scales linearly with the number of APs and remains robust without requiring explicit per-AP channel state information (CSI). Our main contributions are: (i) a multi-receiver Transformer-based decoding model that leverages cross-attention and whose complexity scales linearly with the number of APs, (ii) a fusion mechanism that adapts to heterogeneous link qualities and noise levels across receivers, and (iii) a single trainable decoder that outputs log-likelihood ratios (LLRs) suitable for modern channel decoders.

%A systematic exploration of state-space models for joint 2D time–frequency processing and multi-AP fusion is left for future work.

% The remainder of this paper is organized as follows. Section~\ref{sec:sota} reviews the state of the art. Section~\ref{sec:sys} details the system model. Section~\ref{sec:Transformer} introduces our proposed transformer-based joint decoder. Section~\ref{sec:perf} presents the performance evaluation, and Section~\ref{sec:concl} concludes the paper.

\section{State of the art}\label{sec:sota}

This section reviews receiver designs for point-to-point uplink OFDM, focusing on decoding reliability and complexity under practical constraints such as sparse pilots, non‑stationary channels, and coordinated multi‑AP reception. 
%We first recall model‑based channel estimation/equalization pipelines and then survey data‑driven neural receivers. We finally discuss multi‑AP fusion strategies and the trade-offs between scalability and robustness, highlighting the gap addressed by the proposed cross‑attention Transformer.

\subsection{Classical estimators: LS and LMMSE}

Conventional OFDM receivers estimate the channel from pilots, equalize per subcarrier, and demap to soft information. With a comb or block pilot pattern, the Least Squares (LS) \cite{van_de_beek_channel_1995} estimator computes per‑pilot channel samples by element‑wise normalizing the received pilot symbols with their known transmitted values and then reconstructs the full time–frequency channel by interpolation across subcarriers and OFDM symbols. LS is unbiased and lightweight but noise‑sensitive at low Signal-to-Noise Ratio (SNR) and in interference.

When (approximate) second‑order channel/noise statistics are available, Linear Minimum Mean Square Error (LMMSE) estimation reduces the MSE on pilots and, after interpolation, on the full grid \cite{biguesh_trainingbased_2006}. LMMSE gains, however, hinge on covariance knowledge that is often unavailable, device‑dependent, or quickly outdated in non‑stationary deployments. Moreover, both LS and LMMSE are commonly applied independently per AP, ignoring potential inter‑AP spatial correlations carried by the multi‑receiver observations.

After channel estimation, model‑based equalizers (e.g., Zero-Forcing/MMSE per subcarrier) deliver symbol estimates that are demapped into bit‑wise LLRs for the channel decoder. This modular pipeline remains interpretable and standard‑compliant, but its performance is limited by pilot density, interpolation bias, and the lack of cross‑receiver adaptation in multi‑AP reception.

\subsection{Point-to-point data-driven receivers}

Learned receivers replace some or all model‑based blocks with a neural network trained to output soft information directly from the received resource grid. This paradigm can implicitly learn channel estimation, equalization, interference mitigation, and soft demapping. 

\subsubsection{CNN-based receiver} Convolutional Neural Networks (CNNs) exploit local time–frequency correlations on the 2D OFDM grid; fully convolutional designs learn to denoise, interpolate, equalize, and demap jointly \cite{ye_power_2018,honkala_deeprx_2021}. End-to-end training can reduce pilot overhead without BER loss \cite{ait_aoudia_end--end_2022}. They are parameter‑efficient and accelerator‑friendly but offer limited long‑range context and can be fragile under highly selective fades.

\subsubsection{LSTM-based receiver} Long Short-Term Memory (LSTM) receivers process a sequence of time‑ordered vectors (e.g., per‑subcarrier features per OFDM symbol), maintaining a latent state that tracks channel dynamics and smooths noisy observations \cite{oshea_introduction_2017}, which improves robustness to time selectivity and sparse pilots.

\subsubsection{Transformer-based receiver} Transformers capture long‑range, context‑dependent interactions via attention \cite{vaswani_attention_2017}; on OFDM grids, self‑attention models non‑local dependencies and handles masked resource elements (REs). Attention‑based receivers report robustness and performance gains over MMSE/CNN baselines across diverse multipath profiles through learned positional encodings and context‑aware combining \cite{xie_comm-transformer_2024}.

\subsection{From per‑AP processing to coordinated multi‑AP uplink}

In coordinated architectures (CoMP/cell‑free), geographically diverse observations are exploited to improve reliability \cite{gesbert_multicell_2010,ngo_cellfree_2017}. A practical baseline runs a point‑to‑point chain at each AP and fuses symbols or LLRs centrally (unweighted or SNR/noise‑based), which is simple but not frequency‑selective and ignores inter‑AP correlation; fully joint linear processing can exploit such correlation but demands high‑rate fronthaul, costly inversions, and accurate joint statistics, challenging scalability and real‑time operation \cite{bjornsoN_scalable_2020}.

\subsection{Learned cooperative equalization, FD-RAN architectures}

Recent work has started to explore learned cooperative reception and equalization in cell-free and multi-AP settings. Zecchin \emph{et al.}~\cite{zecchiN_cell-free_2024} propose an in-context learning equalizer for cell-free multi-user MIMO, where a decoder-only Transformer operates on pilot and data observations to adapt to varying channel statistics and fronthaul constraints, outperforming linear MMSE equalization in terms of MSE under pilot contamination and quantized fronthaul. Building on this line, Song \emph{et al.}~\cite{song_-context_2025} investigate state-space models as a more computationally efficient alternative to Transformer-based sequence models for in-context equalization in cell-free massive MIMO, achieving comparable performance with significantly fewer parameters and FLOPs thanks to linear complexity in the context length. In parallel, Zhao \emph{et al.}~\cite{zhao_fully-decoupled_2023} introduce the fully-decoupled RAN (FD-RAN) architecture, which targets resilient uplink cooperative reception via local combining at each base station and centralized combining at the CPU.

Complementary to these algorithmic advances, recent work has explored Spiking Neural Networks (SNNs) for energy-efficient MIMO detection~\cite{song_neuromorphic_2024}. By replacing conventional ANN attention blocks with SNNs, neuromorphic implementations achieve significant power reduction on digital CMOS hardware. While neuromorphic computing addresses hardware efficiency, our contribution focuses on algorithmic design.

While these contributions demonstrate the benefits of cooperative processing and sequence-model-based equalization in cell-free architectures, they typically operate on flat-fading or block-fading MIMO models rather than full OFDM time–frequency grids, and focus on equalization quality instead of producing decoder-ready soft LLRs. Moreover, existing in-context equalizers based on full self-attention scale quadratically with the context length and do not directly address scalable, per-resource-element fusion across a variable number of coordinated APs. Our work is complementary: we target joint multi-AP decoding on 2D OFDM resource grids, with a Transformer architecture that outputs bit-wise LLRs and uses token-wise cross-attention for scalable (linear complexity), robustness-oriented fusion across receivers.

\section{System model and problem formulation}\label{sec:sys}

In this section, we present the system model and problem formulation, and we state the operating assumptions regarding time/frequency synchronization, pilot allocation, and fronthaul characteristics.

\subsection{Assumptions}

\begin{itemize}
  \item A1: Time and frequency synchronization between the UE and APs is either ideal or the residual offsets are within a small bounded range handled by the receiver.
  \item A2: Pilot positions (pilot mask) are fixed and common to all APs, but are not explicitly provided as side information to the neural receivers.
  \item A3: Low-latency, lossless fronthaul (e.g., optical fiber) allowing centralized processing of raw observations $\{\mathbf{Y}^{(r)}\}_{r=1}^{N_R}$, where \(N_R\) is the number of APs.
 \end{itemize}

\subsection{OFDM Transmission Model}

We consider an uplink OFDM transmission scenario where a single-antenna User Equipment (UE) communicates with a set of $N_R$ coordinated APs, each equipped with a single receive antenna. The transmission spans $N_c$ subcarriers and $N_s$ OFDM symbols.

The bitstream $\mathbf{b}\in\{0,1\}^k$ is encoded by $\mathcal{C}(\cdot)$ to produce coded bits $\mathbf{c}\in\{0,1\}^n$. These bits are mapped to complex symbols by the mapper $\mathcal{M}_c(\cdot)$ and arranged on the OFDM resource grid by $\mathcal{M}_{rg}(\cdot)$, yielding:
\begin{equation}
\mathbf{X} = \mathcal{M}_{rg}\!\big(\mathcal{M}_c(\mathcal{C}(\mathbf{b}))\big)\in\mathbb{C}^{N_c\times N_s},
\end{equation}
where $\mathbf{X}$ denotes the transmitted resource grid (subcarrier $\times$ OFDM symbol).

\subsection{Channel Model}
The wireless channel is modeled according to the 3GPP TR 38.901 specifications for Urban Microcell (UMi) environments \cite{noauthor_tr_nodate}. Let $\mathbf{H} \in \mathbb{C}^{N_c \times N_s}$ denote the channel matrix, where each element $h_{f,t}$ represents the channel coefficient at subcarrier $f$ and OFDM symbol $t$.

The received signal matrix $\mathbf{Y} \in \mathbb{C}^{N_c \times N_s}$ is given by:

\begin{equation} \mathbf{Y} = \mathbf{H} \circ \mathbf{X} + \mathbf{N} ,\end{equation}

where:

$\mathbf{X} \in \mathbb{C}^{N_c \times N_s}$ is the transmitted resource grid,
$\mathbf{N} \in \mathbb{C}^{N_c \times N_s}$ is the additive white Gaussian noise matrix with entries $n_{f,t} \sim \mathcal{CN}(0, \sigma_n^2)$, where $\sigma_n^2$ is the noise variance, $\circ$ denotes the Hadamard (element-wise) product.

To enable channel estimation and provide reliable anchors for learning-based receivers, a subset of the resource grid is reserved for known pilot symbols, denoted $\mathbf{X}_p$, which are inserted at predefined time–frequency positions. On these pilot REs, classical methods estimate the corresponding channel coefficients $\mathbf{H}_p$ (e.g., LS/LMMSE) and then interpolate/extrapolate across time and frequency to obtain the full channel matrix $\mathbf{H}$. The same pilots are implicitly exploited by deep learning receivers. They act as trusted anchor points that provide sufficient information for the network to infer and compensate for channel‑induced amplitude/phase distortions over the grid.

\subsection{Multi-AP coordination and decoding objective}

\begin{figure}[t]
  \centering
  \includegraphics[width=0.4\textwidth]{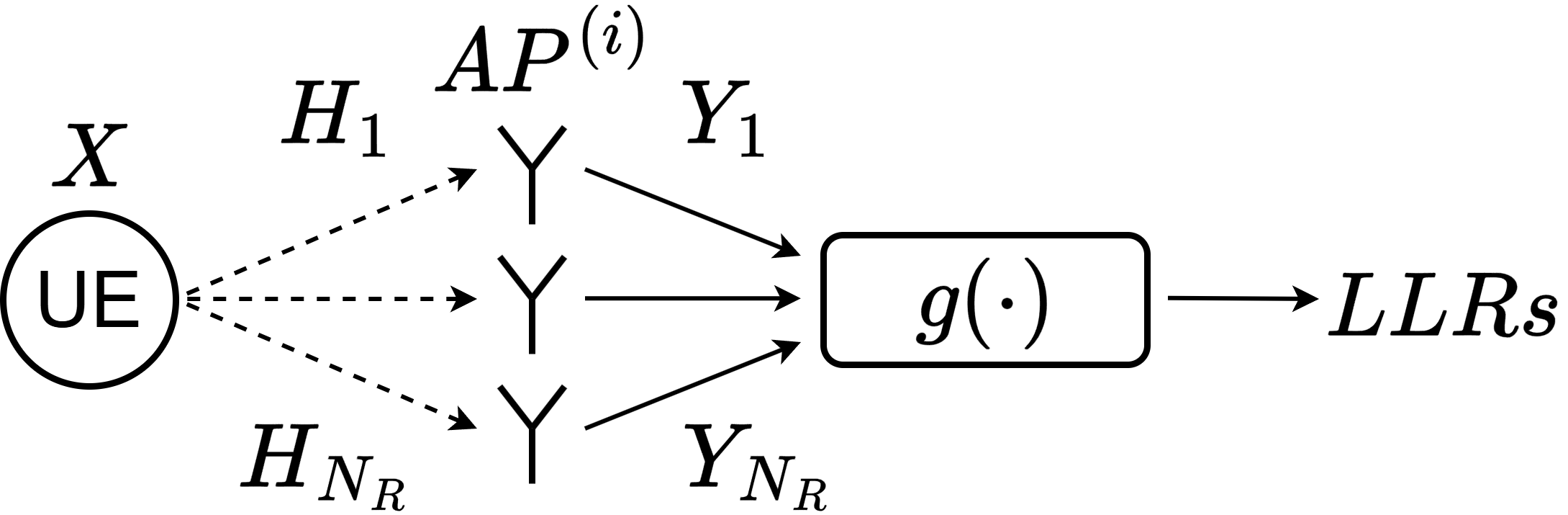}
  \caption{Neural coordinated decoding with three APs.}
  \label{fig:uplink_coord}
\end{figure}

In a coordinated multi-AP uplink scenario, as illustrated in Fig.~\ref{fig:uplink_coord}, a single-antenna UE transmits the signal $\mathbf{X}$ to $N_R$ spatially distributed access points. For the $r$-th AP, the received signal is:
\begin{equation}
    \mathbf{Y}^{(r)} = \mathbf{H}^{(r)} \circ \mathbf{X} + \mathbf{N}^{(r)}, \quad r = 1, \dots, N_R
\end{equation}
where $\mathbf{H}^{(r)}$ denotes the UE‑to‑AP $r$ channel and $\mathbf{N}^{(r)}$ the additive noise at AP $r$ with variance $\sigma_r^2$.

The goal of the neural joint decoder is to process the set of received signals $\{\mathbf{Y}^{(r)}\}_{r=1}^{N_R}$ to produce soft information about the transmitted coded bits $\mathbf{c}$. This is formulated as a function $g_{\theta}$ parameterized by the learnable weights $\theta$, which computes bit-wise log-likelihood ratios (LLRs):
\begin{equation}
  \mathbf{LLR} \;=\; g_{\theta}\!\left(\{\mathbf{Y}^{(r)}\}_{r=1}^{N_R}, \{\sigma_r^2\}_{r=1}^{N_R}\right)
\end{equation}
where $\mathbf{LLR} \in \mathbb{R}^n$ is the vector of LLRs for the $n$ coded bits. The function learns to fuse the multi-AP observations without requiring explicit per-AP channel state information (CSI).

We train the joint decoder $g_{\theta}$ by maximizing the Bit-Metric Decoding (BMD) rate, denoted by $R_{\mathrm{BMD}}$. This objective serves as a differentiable, system-level surrogate for link reliability. In practice, maximizing $R_{\mathrm{BMD}}$ is strongly correlated with minimizing the BER \cite{ait_aoudia_end--end_2022}, whereas the BER itself corresponds to a non-differentiable loss. Letting $s_i \triangleq 2c_i-1 \in \{-1,+1\}$ be the signed transmitted bits, we solve:
\begin{equation}
  \max_{\theta}\; R_{\mathrm{BMD}}(\theta)
  = 1 - \frac{1}{n\ln 2}\,\mathbb{E}_{\mathbf{c}}\!\left[\sum_{i=1}^{n}\log\!\big(1+e^{-s_i L_{i}}\big)\right]
  \label{eq:obj-bmd}
\end{equation}
where the expectation is over the transmitted coded bits $\mathbf{c}$.

Maximizing BMD rate is equivalent (up to a constant) to minimizing the average binary cross-entropy (BCE) on the bits:
\begin{equation}
  R_{\mathrm{BMD}}(\theta) = 1 - \mathcal{L}_{\mathrm{BCE}}(\theta)
  \label{eq:obj-bce}
\end{equation}

Following the neural receiver, the estimated LLRs $\mathbf{LLR}_\theta$ are fed into a standard channel decoder. In our case, a Low-density parity-check (LDPC) decoder is used. The decoder processes this soft information to correct errors and produce the final estimate of the information bits, $\hat{\mathbf{b}}$. This modular approach allows the neural receiver to act as a drop-in replacement for the conventional chain of channel estimation, equalization, and demapping while leveraging the powerful error-correction capabilities of standard channel codes. The end-to-end performance of the system is then evaluated by comparing the decoded bits $\hat{\mathbf{b}}$ against the original transmitted bits $\mathbf{b}$ to compute the BER. For visualization or comparison purposes, an estimate of the transmitted resource grid, $\hat{\mathbf{X}}$, can be reconstructed by re-applying the channel coding and modulation scheme to $\hat{\mathbf{b}}$.

\begin{figure*}[!t]
    \centering
    \includegraphics[width=\textwidth]{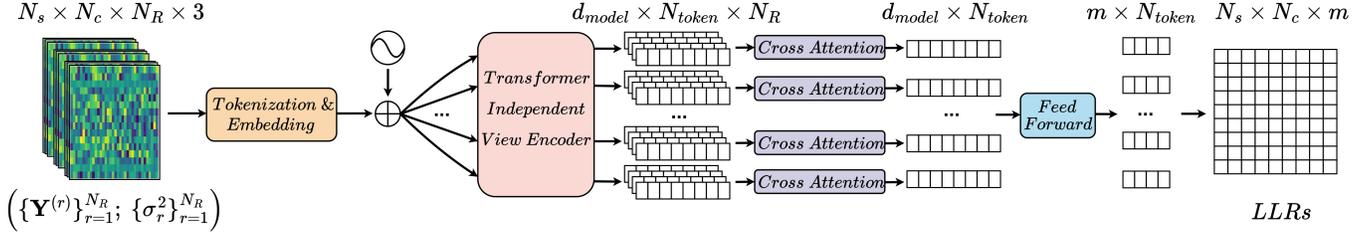}
    \caption{Architecture of the proposed cross-attention Transformer joint decoder.}
    \label{fig:Transformer-arch}
\end{figure*}

\section{Proposed cross-attention Transformer joint decoder}\label{sec:Transformer}

In order to estimate these LLRs, we propose a joint decoder based on a cross-attention Transformer architecture adapted to multi-receiver OFDM signals. The core idea is to first process the received time-frequency (TF) grid from each AP independently with a shared self-attention encoder to extract local features, and then to fuse these features across all APs using a dedicated cross-attention mechanism. This fusion is performed at the granularity of individual REs, allowing the model to adaptively weight each AP signal for each specific time–frequency bin while maintaining a lightweight and computationally efficient architecture.

\subsection{Network architecture}
The overall architecture, depicted in Fig.~\ref{fig:Transformer-arch}, consists of three main stages:
\begin{enumerate}
    \item \textbf{Per-AP shared encoder:} A Transformer encoder with self-attention, shared across all $N_R$ APs, processes the TF grid of each receiver independently. It learns to extract a latent representation for each RE, capturing local and global dependencies.
    \item \textbf{Token-wise cross-attention fusion:} For each TF position $(f,t)$, a cross-attention module fuses the $N_R$ latent representations produced by the encoders. This module learns to dynamically combine information from all APs, effectively up-weighting reliable signals and down-weighting noisy or faded ones.
    \item \textbf{Prediction head:} A simple Multi-Layer Perceptron (MLP) maps the fused representation of each RE to the corresponding bit-level LLRs.
\end{enumerate}
This design enables scalability with $N_R$ and robustness to link failures, as the shared encoder parameters remain constant regardless of $N_R$, and the fusion mechanism can learn to ignore missing or corrupted inputs.

\subsection{Per-AP shared encoder with self-attention}
For each AP $r$, the received complex grid $\mathbf{Y}^{(r)} \in \mathbb{C}^{N_c \times N_s}$ is first transformed into a sequence of input tokens, yielding a total of $N_{token} = N_c\times N_s$ tokens. For each RE at subcarrier $f$ and symbol $t$, we form a vector containing the real and imaginary parts of the received symbol and the estimated noise variance at that AP:
\begin{equation}
    \mathbf{u}^{(r)}_{f,t} = \begin{bmatrix} \operatorname{Re}(Y^{(r)}_{f,t}) & \operatorname{Im}(Y^{(r)}_{f,t}) & \sigma_r^2 \end{bmatrix}^T \in \mathbb{R}^{3}.
\end{equation}

These vectors are treated as $1 \times 1$ patches. Each token is linearly projected into the model latent dimension $d_{\text{model}}$ and augmented with a 2D sinusoidal positional encoding $\pi_{f,t}$ to retain its TF position information:
\begin{equation}
    \mathbf{z}^{(r)}_{0,f,t} = W_e \mathbf{u}^{(r)}_{f,t} + \pi_{f,t} \in \mathbb{R}^{d_{\text{model}}},
\end{equation}
where $W_e$ is a shared embedding matrix.

The resulting sequence of $N_{token}$ tokens for AP $r$, denoted $\mathbf{Z}^{(r)}_0$, is fed into a stack of 4 encoder layers. Each layer applies multi-head self-attention (MHSA) to capture dependencies across the entire TF grid. For a given sequence of input embeddings $\mathbf{Z}$, the scaled dot-product attention is defined as:
\begin{equation}
\label{eq:self-attention}
    \text{Attention}(Q, K, V) = \text{softmax}\left(\frac{QK^T}{\sqrt{d_k}}\right)V,
\end{equation}
where the queries $Q$, keys $K$, and values $V$ are linear projections of the input sequence $\mathbf{Z}$ (i.e., $Q=\mathbf{Z}W_Q, K=\mathbf{Z}W_K, V=\mathbf{Z}W_V$). The self-attention mechanism allows the model to learn context-aware representations for each RE by attending to all other REs in the same grid.

\subsection{Token-wise anchor-query cross-attention}

After the shared per-AP encoder, we perform fusion for each TF position $(f,t)$ independently. For a given position $(f,t)$, we consider the sequence of $N_R$ output embeddings from the encoders, one for each AP:
\begin{equation}
    \mathbf{Z}_{f,t} = (\mathbf{z}^{(1)}_{f,t}, \mathbf{z}^{(2)}_{f,t}, \dots, \mathbf{z}^{(N_R)}_{f,t}) \in \mathbb{R}^{d_{\text{model}} \times N_R}.
\end{equation}
This sequence is treated as a set of $N_R$ tokens, each of dimension $d_{\text{model}}$.

% Fusion is performed using an anchor-based cross-attention mechanism. We designate AP 1 as the "anchor" without loss of generality (any AP could serve this role), while all views contribute to the keys and values. The query $\mathbf{q}_{f,t}$, keys $\mathbf{K}_{f,t}$, and values $\mathbf{V}_{f,t}$ are computed as follows:
Fusion is performed using an anchor-based cross-attention mechanism. The AP order is randomly assigned and so AP 1 is used only as an arbitrary reference. All APs contribute to the keys and values. The query $\mathbf{q}_{f,t}$, keys $\mathbf{K}_{f,t}$, and values $\mathbf{V}_{f,t}$ are computed as follows:

\begin{align}
    \mathbf{q}_{f,t} &= \mathbf{z}^{(1)}_{f,t} W_Q \in \mathbb{R}^{1 \times d_k}, \\
    \mathbf{K}_{f,t} &= \mathbf{Z}_{f,t} W_K \in \mathbb{R}^{N_R \times d_k}, \\
    \mathbf{V}_{f,t} &= \mathbf{Z}_{f,t} W_V \in \mathbb{R}^{N_R \times d_v},
\end{align}
where $W_Q$, $W_K$, and $W_V$ are learnable projection matrices, and we assume the sequence $\mathbf{Z}_{f,t}$ is formatted as a matrix of size $N_R \times d_{\text{model}}$. The attention output $\mathbf{a}_{f,t}$ is a weighted sum of the values:
\begin{equation}
  \mathbf{a}_{f,t} = \text{softmax}\left(\frac{\mathbf{q}_{f,t}\mathbf{K}_{f,t}^T}{\sqrt{d_k}}\right)\mathbf{V}_{f,t} \in \mathbb{R}^{1 \times d_v}.
\end{equation}
We then apply a residual connection to the anchor embedding, followed by layer normalization, to obtain the fused representation:
\begin{equation}
  \mathbf{z}^{\text{fused}}_{f,t} = \mathrm{LN}\big(\mathbf{z}^{(1)}_{f,t} + \mathbf{a}_{f,t}\big) \in \mathbb{R}^{d_{\text{model}}}
\end{equation}
A lightweight MLP finally maps $\mathbf{z}^{\text{fused}}_{f,t}$ to $m$ logits (bit LLRs) per RE, where $m$ is the number of bits per QAM symbol.

\begin{equation}
    \mathbf{LLR}_{f,t} = \text{MLP}(\mathbf{z}^{\text{fused}}_{f,t}) \in \mathbb{R}^{m}.
\end{equation}

% \textbf{Remark:} The anchor choice (AP 1) is arbitrary and used only to 
% define the query vector. The fusion weights depend on all AP embeddings and can down-weight unreliable views. 

% Compared to large-sequence designs that perform full self-attention over concatenated per-AP sequences \cite{yu_large_2025}, our token-wise cross-attention restricts the fusion to the $N_R$ views corresponding to the same time–frequency position. This preserves the 2D OFDM structure and scales linearly in $N_R$ per RE.

Compared to large-sequence designs performing full self-attention over concatenated per-AP sequences \cite{yu_large_2025}, our approach focuses complexity on the fusion module rather than global attention across the multi-AP sequence. The proposed token-wise cross-attention restricts fusion to the $N_R$ views corresponding to the same TF position, thereby preserving the 2D OFDM structure. As a result, fusion complexity scales linearly with the number of receivers $N_R$ and the OFDM grid size $(N_c \times N_s)$. The overall model complexity remains quadratic in the OFDM grid dimensions $(N_c \times N_s)$, primarily due to the per-AP self-attention encoders.

\section{Performance evaluation}\label{sec:perf}

\subsection{Simulation setup}

We evaluate the proposed joint decoder against several baselines: (i) classical LS and LMMSE pipelines, (ii) a CNN-based receiver from \cite{ait_aoudia_end--end_2022}, (iii) a full self-attention fusion Transformer baseline inspired by \cite{yu_large_2025} and adapted to our 2D OFDM resource grid. This model is a refined version of the cell-free in-context learning equalizer of \cite{zecchiN_cell-free_2024}. However, the original cell-free architectures cannot be directly applied here due to the much larger problem dimension, which would require full self-attention over all TF–AP tokens, and (iv) a global MMSE receiver with perfect CSI, used as an upper-bound reference.

In the multi-AP case, the LS, LMMSE, and CNN baselines first generate LLRs independently at each AP and are then fused centrally using SNR-based weighting. By contrast, the perfect-CSI upper bound performs joint global MMSE combining across APs and therefore does not rely on post-hoc LLR fusion. The full-attention baseline mirrors our architecture at the per-AP level (shared Transformer encoder on each received grid) but replaces the token-wise cross-attention fusion by a multi-head self-attention layer applied jointly to the concatenated per-AP tokens, in the spirit of \cite{yu_large_2025}.

All simulations use the 3GPP TR 38.901 Urban Microcell (UMi) channel
model to capture realistic multipath fading and user mobility. Key parameters are summarized in
Table~\ref{tab:params}.

\begin{table}[!h]
\centering
\caption{Simulation Parameters}
\label{tab:params}
\renewcommand{\arraystretch}{1.1} % Adds a bit of vertical space
\begin{tabular}{ll}
\toprule
\textbf{Parameter} & \textbf{Value} \\
\midrule
Carrier Frequency & 2.4 GHz \\
Bandwidth & 20 MHz \\
Subcarrier Spacing & 15 kHz \\
FFT Size & 1024 \\
Number of Subcarriers ($N_c$) & 48 \\
Number of OFDM Symbols ($N_s$) & 36 \\
Modulation ($m=6$) & 64-QAM \\
Channel Coding & LDPC 3/4\\
Channel Model & 3GPP TR 38.901 UMi \\
UE Speed & 0-3 m/s \\
Number of APs ($N_R$) & 1, 3, 10 \\
\bottomrule
\end{tabular}
\end{table}

\subsection{Data generation}

To ensure the model generalizes across diverse channel conditions and avoids overfitting, both training and evaluation data are generated on-the-fly. For each sample, a new scenario is created by randomly placing the single-antenna UE and the $N_R$ single-antenna APs within a 25m $\times$ 25m square area. The entire simulation pipeline is implemented using the \textbf{Sionna} library \cite{sionna}, which provides tools for link-level simulation.
\subsection{Experimental Protocol}

\textbf{Training:} Our model is trained for 30,000 steps using the Adam optimizer. A batch size of 16 is used, where each item in the batch corresponds to a full multi-AP observation $\{\mathbf{Y}^{(r)}\}_{r=1}^{N_R}$ from an independently generated random topology.

\textbf{Evaluation:} We compute the BER using 5,000 Monte Carlo iterations. In every iteration, a random UE/AP placement is generated, and a batch of 16 independent resource grids is transmitted. Each iteration is evaluated at the mean $E_b/N_0$ across the $N_R$ receive links and yields one BER sample at that $E_b/N_0$. To reduce run-to-run variability, we repeat the entire evaluation 5 times with independent random seeds and report the mean BER across the five runs. The final BER curve is then smoothed using kernel smoothing with a 1\,dB bandwidth.

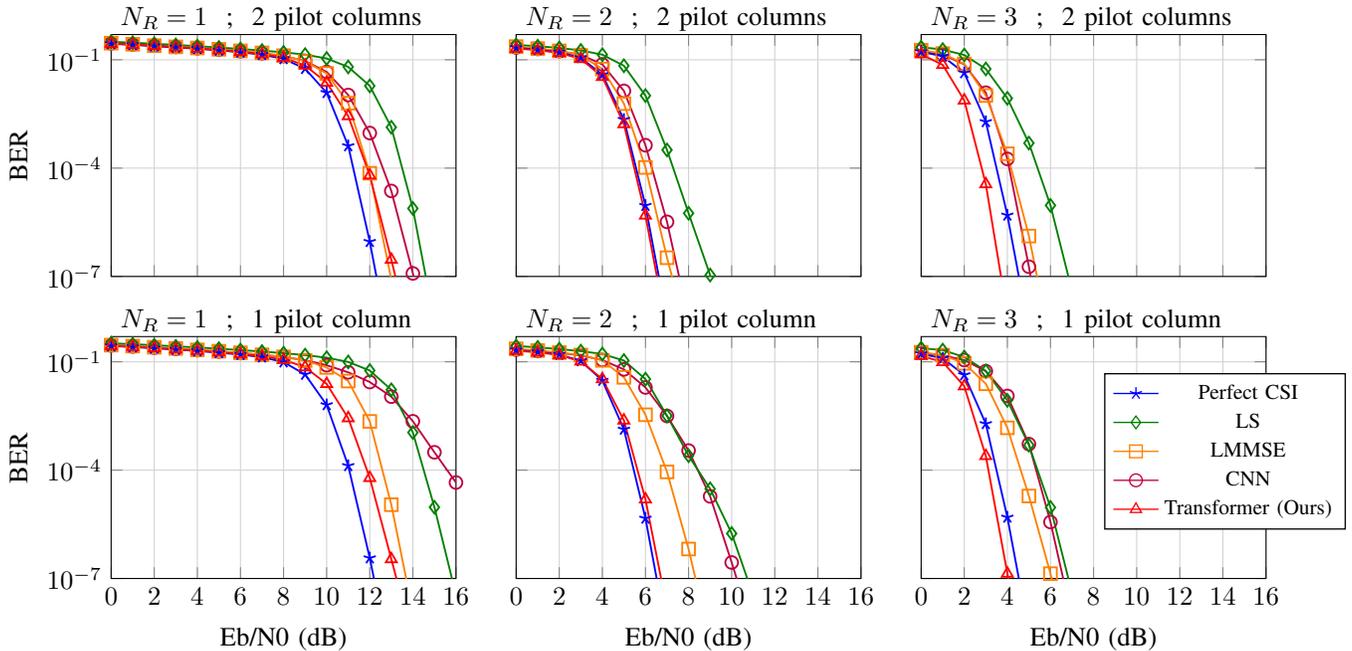
\begin{figure*}[t]
\centering
\begin{tikzpicture}
\def\eps{1e-12} 

\begin{groupplot}[
  group style={group size=3 by 2, horizontal sep=0.7cm, vertical sep=1.2cm},
  width=0.36\textwidth, height=4.4cm,
  % globalement on retire xlabel et ylabel ici
  ymode=log, ymin=1e-7, ymax=5e-1,
  ytick={1e-7, 1e-6,1e-5,1e-4,1e-3,1e-2,1e-1},
  grid=both,
  minor x tick num=1,
  minor y tick num=0,
  major grid style={gray!55}, minor grid style={gray!15},
  every axis plot/.append style={line width=0.7pt, mark options={scale=1.2}},
  title style={font=\normalsize, at={(0,1)}, anchor=west}
]

% Top-left: affiche ylabel uniquement ici, et cache les labels x (top row)
\nextgroupplot[
  title={\(N_R = 1\)  \ ; \  2 pilot columns},
  ylabel={BER},
  xmin=8, xmax=16,
]
% ... \addplot ... (mêmes données)
\addplot+[color=purple, mark=o] coordinates {
    (-1.0, 2.990747e-01) (0.0, 2.842962e-01) (1.0, 2.665070e-01) (2.0, 2.470951e-01)
    (3.0, 2.272759e-01) (4.0, 2.087907e-01) (5.0, 1.913314e-01) (6.0, 1.730011e-01)
    (7.0, 1.516382e-01) (8.0, 1.264664e-01) (9.0, 9.293763e-02) (10.0, 4.613326e-02)
    (11.0, 1.058599e-02) (12.0, 9.364287e-04) (13.0, 2.331557e-05) (14.0, 1.213186e-07)
    (15.0, 3.385362e-10) (16.0, 3.375315e-12) (17.0, 2.634693e-13)
}; 
% LMMSE
\addplot+[color=orange, mark=square] coordinates {
    (-1.0, 3.015576e-01) (0.0, 2.888858e-01) (1.0, 2.692926e-01) (2.0, 2.465400e-01)
    (3.0, 2.280278e-01) (4.0, 2.125958e-01) (5.0, 1.954926e-01) (6.0, 1.763431e-01)
    (7.0, 1.554448e-01) (8.0, 1.299704e-01) (9.0, 9.091928e-02) (10.0, 4.183605e-02)
    (11.0, 6.293717e-03) (12.0, 7.215618e-05) (13.0, 8.166687e-08) (14.0, 1.461557e-10)
    (15.0, 3.841065e-12) (16.0, 1.042497e-12) (17.0, 2.530247e-13)
}; 
% LS
\addplot+[color=green!50!black, mark=diamond] coordinates {
    (-1.0, 3.297377e-01) (0.0, 3.154390e-01) (1.0, 2.974457e-01) (2.0, 2.779685e-01)
    (3.0, 2.578149e-01) (4.0, 2.382866e-01) (5.0, 2.192097e-01) (6.0, 1.994411e-01)
    (7.0, 1.799513e-01) (8.0, 1.609114e-01) (9.0, 1.389267e-01) (10.0, 1.086773e-01)
    (11.0, 6.340547e-02) (12.0, 1.893542e-02) (13.0, 1.353325e-03) (14.0, 7.646936e-06)
    (15.0, 5.437974e-09) (16.0, 1.205827e-11) (17.0, 4.426961e-13)
};
% PERFECT-CSI
\addplot+[color=blue, mark=star] coordinates {
    (-1.0, 2.919870e-01) (0.0, 2.774060e-01) (1.0, 2.589462e-01) (2.0, 2.397756e-01)
    (3.0, 2.209769e-01) (4.0, 2.021128e-01) (5.0, 1.823475e-01) (6.0, 1.617012e-01)
    (7.0, 1.384509e-01) (8.0, 1.061494e-01) (9.0, 5.663392e-02) (10.0, 1.223025e-02)
    (11.0, 4.121327e-04) (12.0, 9.188626e-07) (13.0, 7.097961e-10) (14.0, 2.292373e-12)
    (15.0, 7.176390e-14) (16.0, 9.401379e-15) (17.0, 2.961390e-15)
}; 

\addplot+[color=magenta, mark=x] coordinates {
    (-1.0, 2.942566e-01)
    (0.0, 2.795865e-01)
    (1.0, 2.609990e-01)
    (2.0, 2.420415e-01)
    (3.0, 2.230280e-01)
    (4.0, 2.037273e-01)
    (5.0, 1.847844e-01)
    (6.0, 1.654274e-01)
    (7.0, 1.434189e-01)
    (8.0, 1.038999e-01)
    (9.0, 6.593204e-02)
    (10.0, 1.845663e-02)
    (11.0, 1.150615e-03)
    (12.0, 2.193188e-05)
    (13.0, 7.444399e-08)
};
\addplot+[color=red, mark=triangle, solid] coordinates {
    (-1.0, 2.962947e-01) (0.0, 2.810651e-01) (1.0, 2.628711e-01) (2.0, 2.430114e-01)
    (3.0, 2.235173e-01) (4.0, 2.052493e-01) (5.0, 1.866256e-01) (6.0, 1.667722e-01)
    (7.0, 1.443317e-01) (8.0, 1.146398e-01) (9.0, 6.937843e-02) (10.0, 2.324335e-02)
    (11.0, 2.759269e-03) (12.0, 6.323792e-05) (13.0, 2.929871e-07) (14.0, 1.004820e-09)
    (15.0, 9.257854e-12) (16.0, 3.304827e-13) (17.0, 4.528144e-14)
}; 

% Top-right
\nextgroupplot[
  title={\(N_R = 3\)  \ ; \  2 pilot columns},
  yticklabels={}, 
  xmin=0, xmax=8,
]
% ... \addplot ...
\addplot+[color=purple, mark=o] coordinates {
  (-1.0,1.989e-01) (0.0,1.762e-01) (1.0,1.393e-01) (2.0,7.450e-02)
  (3.0,1.225e-02) (4.0,1.799e-04) (5.0,1.856e-07) (6.0,6.546e-11)
  (7.0,1.104e-13) (8.0,5.045e-15) (9.0,1.809e-15) (10.0,1.281e-15)
  (11.0,1.131e-15) (12.0,1.072e-15)
};
\addplot+[color=orange, mark=square] coordinates {
  (-1.0,2.085e-01) (0.0,1.848e-01) (1.0,1.485e-01) (2.0,7.652e-02)
  (3.0,1.024e-02) (4.0,2.500e-04) (5.0,1.315e-06) (6.0,1.391e-09)
  (7.0,1.984e-12) (8.0,2.056e-14) (9.0,2.159e-15) (10.0,1.117e-15)
  (11.0,1.008e-15) (12.0,1.000e-15)
};
\addplot+[color=green!50!black, mark=diamond] coordinates {
  (-1.0,2.455e-01) (0.0,2.233e-01) (1.0,1.916e-01) (2.0,1.357e-01)
  (3.0,5.577e-02) (4.0,8.561e-03) (5.0,4.927e-04) (6.0,9.275e-06)
  (7.0,3.971e-08) (8.0,9.172e-11) (9.0,3.909e-13) (10.0,1.021e-14)
  (11.0,1.871e-15) (12.0,1.118e-15)
};
\addplot+[color=blue, mark=star] coordinates {
(-1.0,1.473e-01)
(0.0,1.123e-01)
(1.0,4.100e-02)
(2.0,2.579e-03)
(3.0,5.130e-06)
(4.0,5.283e-10)
};
\addplot+[color=magenta, mark=x] coordinates {
  (-1.0, 1.80319e-01)
  (0.0, 1.44516e-01)
  (1.0, 9.4655e-02)
  (2.0, 1.7396e-02)
  (3.0, 4.1702e-04)
  (4.0, 9.4710e-07)
  (5.0, 5.2270e-10)
};
\addplot+[color=red, mark=triangle, solid] coordinates {
    (-1.0, 1.724381e-01)
    (0.0, 1.424318e-01)
    (1.0, 7.240992e-02)
    (2.0, 7.501164e-03)
    (3.0, 3.604633e-05)
    (4.0, 8.532872e-09)
    (5.0, 2.578257e-12)
    (6.0, 1.998143e-14)
    (7.0, 2.678138e-15)
    (8.0, 1.400851e-15)
    (9.0, 1.124709e-15)
};

% Top-middle: pas de ylabel, pas de xticklabels
\nextgroupplot[
  title={\(N_R = 10\)  \ ; \  2 pilot columns},
  yticklabels={}, 
  xmin=-8,xmax=0
]

% blue = baseline-perfect-csi-joint
\addplot+[color=blue, mark=star] coordinates {
  (-8.0,1.3e-01)
  (-7.0,4e-02)
  (-6.0,2.0e-03)
  (-5.0,1.5e-05)
  (-4.0,5.0e-09)
  (-3.0,5.0e-11)
  (-2.0,1.2e-13)
  (-1.0,2.0e-15)
  (0.0,1.0e-15)
  (1.0,1.0e-15)
  (2.0,1.0e-15)
  (3.0,1.0e-15)
  (4.0,1.0e-15)
  (5.0,1.0e-15)
  (6.0,1.0e-15)
  (7.0,1.0e-15)
};

% orange = baseline-ls-averaged
\addplot+[color=green!50!black, mark=diamond] coordinates {
  (-8.0,2.1e-01)
  (-7.0,1.9e-01)
  (-6.0,1.7e-01)
  (-5.0,1.5e-01)
  (-4.0,1.3e-01)
  (-3.0,1.1e-01)
  (-2.0,9.5e-02)
  (-1.0,8.0e-02)
  (0.0,6.8e-02)
  (1.0,5.8e-02)
  (2.0,4.8e-02)
  (3.0,3.8e-02)
  (4.0,2.7e-02)
  (5.0,1.5e-02)
  (6.0,8.0e-03)
};

% green = baseline-lmmse-averaged
\addplot+[color=orange, mark=square] coordinates {
  (-8.0,2.0e-01)
  (-7.0,1.6e-01)
  (-6.0,1.4e-01)
  (-5.0,8.0e-02)
  (-4.0,2.5e-02)
  (-3.0,8.0e-03)
  (-2.0,2.5e-03)
  (-1.0,6.0e-04)
  (0.0,8.0e-05)
  (1.0,2.0e-06)
  (2.0,4.0e-08)
  (3.0,1.5e-09)
  (4.0,1.2e-10)
  (5.0,1.0e-11)
};

% red = neural-receiver-transformer
\addplot+[color=red, mark=triangle, solid] coordinates {
  (-8.0,1.5e-01)
  (-7.0,8e-02)
  (-6.0,1.0e-02)
  (-5.0,1.4e-04)
  (-4.0,1.2e-07)
  (-3.0,3.0e-10)
  (-2.0,2.5e-13)
  (-1.0,1.2e-14)
  (0.0,2.0e-15)
  (1.0,1.0e-15)
  (2.0,1.0e-15)
  (3.0,1.0e-15)
  (4.0,1.0e-15)
  (5.0,1.0e-15)
  (6.0,1.0e-15)
  (7.0,1.0e-15)
};

% magenta = neural-receiver-cnn
\addplot+[color=purple, mark=o] coordinates {
  (-8.0,2.0e-01)
  (-7.0,1.6e-01)
  (-6.0,9.0e-02)
  (-5.0,2.5e-02)
  (-4.0,6.0e-04)
  (-3.0,3.0e-06)
  (-2.0,2.0e-09)
  (-1.0,4.0e-12)
  (0.0,5.0e-14)
  (1.0,3.0e-15)
  (2.0,1.2e-15)
  (3.0,1.0e-15)
  (4.0,1.0e-15)
  (5.0,1.0e-15)
  (6.0,1.0e-15)
  (7.0,1.0e-15)
};

% Bottom-left: afficher ylabel (déjà répété), ici on veut garder les xticks visibles
\nextgroupplot[
  title={\(N_R = 1\)  \ ; \  1 pilot column},
  ylabel={BER}, % seul la colonne de gauche a BER comme label
  xlabel={Eb/N0 (dB)},
  xmin=8, xmax=16,
]
% ... \addplot ...
% CNN
\addplot+[color=purple, mark=o] coordinates {
    (-1.0, 2.984883e-01) (0.0, 2.845766e-01) (1.0, 2.662144e-01) (2.0, 2.465856e-01)
    (3.0, 2.280760e-01) (4.0, 2.105887e-01) (5.0, 1.933462e-01) (6.0, 1.748303e-01)
    (7.0, 1.561509e-01) (8.0, 1.362703e-01) (9.0, 1.104731e-01) (10.0, 8.051846e-02)
    (11.0, 5.119192e-02) (12.0, 2.753846e-02) (13.0, 1.082941e-02) (14.0, 2.258561e-03)
    (15.0, 3.126762e-04) (16.0, 4.543460e-05) (17.0, 9.484502e-06)
};
% LMMSE
\addplot+[color=orange, mark=square] coordinates {
    (-1.0, 3.023109e-01) (0.0, 2.899933e-01) (1.0, 2.724903e-01) (2.0, 2.529750e-01)
    (3.0, 2.346441e-01) (4.0, 2.142676e-01) (5.0, 1.960149e-01) (6.0, 1.777105e-01)
    (7.0, 1.566077e-01) (8.0, 1.372307e-01) (9.0, 1.087666e-01) (10.0, 6.980842e-02)
    (11.0, 2.936799e-02) (12.0, 2.235785e-03) (13.0, 1.103562e-05) (14.0, 1.187646e-08)
    (15.0, 7.982886e-11) (16.0, 7.343327e-12) (17.0, 6.948130e-13)
};
% LS
\addplot+[color=green!50!black, mark=diamond] coordinates {
    (-1.0, 3.403906e-01) (0.0, 3.260892e-01) (1.0, 3.078468e-01) (2.0, 2.880429e-01)
    (3.0, 2.686083e-01) (4.0, 2.497326e-01) (5.0, 2.307521e-01) (6.0, 2.115486e-01)
    (7.0, 1.927629e-01) (8.0, 1.734302e-01) (9.0, 1.532911e-01) (10.0, 1.301345e-01)
    (11.0, 9.911297e-02) (12.0, 5.737772e-02) (13.0, 1.680188e-02) (14.0, 1.101183e-03)
    (15.0, 9.402653e-06) (16.0, 3.522756e-08) (17.0, 5.187036e-10)
};
% PERFECT-CSI
\addplot+[color=blue, mark=star] coordinates {
    (-1.0, 2.887559e-01) (0.0, 2.743891e-01) (1.0, 2.563654e-01) (2.0, 2.370533e-01)
    (3.0, 2.185581e-01) (4.0, 1.995075e-01) (5.0, 1.790661e-01) (6.0, 1.586279e-01)
    (7.0, 1.347245e-01) (8.0, 9.831821e-02) (9.0, 4.461667e-02) (10.0, 6.399168e-03)
    (11.0, 1.322210e-04) (12.0, 3.677022e-07) (13.0, 4.344659e-10) (14.0, 1.427005e-12)
    (15.0, 3.705575e-14) (16.0, 4.922093e-15) (17.0, 1.937221e-15)
};
\addplot+[color=magenta, mark=x] coordinates {
    (-1.0, 2.942566e-01)
    (0.0, 2.795865e-01)
    (1.0, 2.609990e-01)
    (2.0, 2.420415e-01)
    (3.0, 2.230280e-01)
    (4.0, 2.037273e-01)
    (5.0, 1.847844e-01)
    (6.0, 1.654274e-01)
    (7.0, 1.434189e-01)
    (8.0, 1.038999e-01)
    (9.0, 6.593204e-02)
    (10.0, 1.845663e-02)
    (11.0, 1.150615e-03)
    (12.0, 2.193188e-05)
    (13.0, 7.444399e-08)
    (14.0, 1.942189e-09)
    (15.0, 3.328812e-11)
    (16.0, 1.179075e-12)
    (17.0, 1.384166e-13)
};
% TRANSFORMER
\addplot+[color=red, mark=triangle,solid] coordinates {
    (-1.0, 2.942566e-01)
    (0.0, 2.795865e-01)
    (1.0, 2.609990e-01)
    (2.0, 2.420415e-01)
    (3.0, 2.230280e-01)
    (4.0, 2.037273e-01)
    (5.0, 1.847844e-01)
    (6.0, 1.654274e-01)
    (7.0, 1.434189e-01)
    (8.0, 1.138999e-01)
    (9.0, 7.093204e-02)
    (10.0, 2.445663e-02)
    (11.0, 2.750615e-03)
    (12.0, 6.093188e-05)
    (13.0, 3.444399e-07)
    (14.0, 1.942189e-09)
    (15.0, 3.328812e-11)
    (16.0, 1.179075e-12)
    (17.0, 1.384166e-13)
};

% Bottom-right
\nextgroupplot[
  title={\(N_R = 3\)  \ ; \  1 pilot column},
  yticklabels={},
  xlabel={Eb/N0 (dB)},
  xmin=0, xmax=8,
]
% ... \addplot ...
\addplot+[color=purple, mark=o] coordinates {
  (-1.0,1.935e-01) (0.0,1.752e-01) (1.0,1.508e-01) (2.0,1.118e-01)
  (3.0,5.505e-02) (4.0,1.135e-02) (5.0,5.275e-04) (6.0,3.655e-06)
  (7.0,8.132e-09) (8.0,3.128e-11) (9.0,4.629e-13) (10.0,2.798e-14)
  (11.0,5.238e-15) (12.0,2.240e-15)
};
\addplot+[color=orange, mark=square] coordinates {
  (0.0,1.841e-01) (1.0,1.527e-01) (2.0,9.088e-02) (3.0,2.466e-02)
  (4.0,1.484e-03) (5.0,1.923e-05) (6.0,1.360e-07) (7.0,4.324e-10)
  (8.0,7.102e-13) (9.0,8.175e-15) (10.0,2.257e-15) (11.0,1.973e-15) (12.0,1.572e-15)
};
\addplot+[color=green!50!black, mark=diamond] coordinates {
  (-1.0,2.580e-01) (0.0,2.390e-01) (1.0,2.111e-01) (2.0,1.357e-01)
  (3.0,5.577e-02) (4.0,8.561e-03) (5.0,4.927e-04) (6.0,9.275e-06)
  (7.0,3.971e-08) (8.0,9.172e-11) (9.0,3.909e-13) (10.0,1.021e-14) (11.0,1.871e-15) (12.0,1.118e-15)
};
\addplot+[color=blue, mark=star] coordinates {
(-1.0,1.473e-01)
(0.0,1.123e-01)
(1.0,4.100e-02)
(2.0,2.579e-03)
(3.0,5.130e-06)
(4.0,5.283e-10)
};

\addplot+[color=magenta, mark=x] coordinates {
  (-1.0,1.912e-01) (0.0,1.666e-01) (1.0,1.449e-01) (2.0,6.329e-02)
  (3.0,3.918e-03) (4.0,2.934e-05) (5.0,2.252e-08) (6.0,4.064e-12)
  (7.0,2.677e-14) (8.0,2.272e-15) (9.0,1.161e-15) (10.0,1.024e-15) (11.0,1.003e-15) (12.0,1.000e-15)
};

\addplot+[color=red, mark=triangle,solid] coordinates {
  (-1.0,1.676e-01) (0.0,1.462e-01) (1.0,9.857e-02) (2.0,2.113e-02)
  (3.0,2.471e-04) (4.0,1.369e-07) (5.0,4.989e-11) (6.0,1.463e-13)
  (7.0,6.977e-15) (8.0,2.161e-15) (9.0,1.379e-15) (10.0,1.128e-15) (11.0,1.051e-15) (12.0,1.041e-15)
};

% Bottom-middle: cacher y-ticks (pas de label vertical), garder xticks visibles
\nextgroupplot[
  title={\(N_R = 10\)  \ ; \  1 pilot column},
  yticklabels={},
  xlabel={Eb/N0 (dB)},
  xmin=-8,xmax=0
]
% blue = baseline-perfect-csi-joint
\addplot+[color=blue, mark=star] coordinates {
  (-8.0,1.3e-01)
  (-7.0,4e-02)
  (-6.0,2.0e-03)
  (-5.0,1.5e-05)
  (-4.0,5.0e-09)
  (-3.0,5.0e-11)
  (-2.0,1.2e-13)
  (-1.0,2.0e-15)
  (0.0,1.0e-15)
  (1.0,1.0e-15)
  (2.0,1.0e-15)
  (3.0,1.0e-15)
  (4.0,1.0e-15)
  (5.0,1.0e-15)
  (6.0,1.0e-15)
  (7.0,1.0e-15)
};

% orange = baseline-ls-averaged
\addplot+[color=green!50!black, mark=diamond] coordinates {
  (-8.0,2.1e-01)
  (-7.0,1.9e-01)
  (-6.0,1.7e-01)
  (-5.0,1.6e-01)
  (-4.0,1.4e-01)
  (-3.0,1.2e-01)
  (-2.0,1.1e-01)
  (-1.0,1.05e-01)
  (0.0,8.8e-02)
  (1.0,6.8e-02)
};

% green = baseline-lmmse-averaged
\addplot+[color=orange, mark=square] coordinates {
(-8.0,2.0e-01)
(-7.0,1.75e-01)
(-6.0,1.40e-01)
(-5.0,9.8e-02)
(-4.0,5.8e-02)
(-3.0,2.5e-02)
(-2.0,8.0e-03)
(-1.0,2.0e-03)
( 0.0,5.0e-04)
};

% red = neural-receiver-transformer
\addplot+[color=red, mark=triangle, solid] coordinates {
(-8.0,1.5e-01)
(-7.0,8.0e-02)
(-6.0,1.2e-02)
(-5.0,2.0e-04)
(-4.0,2.0e-07)
(-3.0,2.0e-11)
};

% magenta = neural-receiver-cnn
\addplot+[color=purple, mark=o] coordinates {
(-8.0,2.0e-01)
(-7.0,1.6e-01)
(-6.0,9.5e-02)
(-5.0,3.2e-02)
(-4.0,1.2e-03)
(-3.0,1.2e-05)
(-2.0,8.0e-09)
};

\end{groupplot}

\node[overlay,below=1.0cm of group c2r2.south, anchor=north] (legend) {
    \begin{tikzpicture}
    \begin{axis}[
        hide axis,
        xmin=0, xmax=1,
        ymin=0, ymax=1,
        legend style={
            draw=black,
            line width=0.5pt,
            fill=white,
            legend columns=6,
            /tikz/every even column/.append style={column sep=0.2cm},
            mark options={scale=1.2}
        }
    ]

    \addlegendimage{color=blue, mark=star}
    \addlegendentry{\footnotesize Global MMSE (perfect CSI)}
    \addlegendimage{color=green!50!black, mark=diamond}
    \addlegendentry{\footnotesize LS}
    \addlegendimage{color=orange, mark=square}
    \addlegendentry{\footnotesize LMMSE}
    \addlegendimage{color=purple, mark=o}
    \addlegendentry{\footnotesize CNN\cite{ait_aoudia_end--end_2022}}
    \addlegendimage{color=magenta, mark=x}
    \addlegendentry{\footnotesize Self-Attn Transf.\cite{yu_large_2025}}
    \addlegendimage{color=red, mark=triangle}
    \addlegendentry{\footnotesize Cross-Attn Transf. (Ours)}

    \end{axis}
    \end{tikzpicture}
};

\end{tikzpicture}
\vspace{0.6cm}
\caption{BER performance vs. Eb/N0 for varying cooperation levels ($N_R = 1, 3, 10$) and pilot configurations (1 vs. 2 pilot columns)}
\label{fig:ber-grid}
\end{figure*}

\subsection{Hyperparameters}
All Transformer blocks (shared encoder and cross-attention fusion) use $d_{\text{model}}=64$, 8 heads, 4 layers, a feed-forward network dimension of 128, and a patch size of $1\times 1$ (per RE). The full self-attention Transformer baseline is configured with the same hyperparameters to ensure a fair comparison. These values were determined through ablation studies across different numbers of heads, layers, and model dimensions.

\subsection{Results and analysis}

\subsubsection{BER performance}

Results are shown in Fig.~\ref{fig:ber-grid} as BER versus the average $E_b/N_0$ across the $N_R$ receive links. We study both the impact of cooperation, with $N_R\in\{1,3,10\}$, and the robustness to pilot sparsity using one or two pilot columns. As expected, increasing the number of coordinated APs improves all receivers through spatial diversity, with BER curves shifting left as $N_R$ increases. The full self-attention baseline is not reported for $N_R=10$, as joint attention over all TF-AP tokens was prohibitively expensive and could not be executed within the available hardware budget, even on an NVIDIA H100 GPU with 80\,GB of memory.

This gain is particularly pronounced for the proposed cross-attention Transformer. For $N_R=3$ with two pilot columns, it outperforms the full self-attention Transformer by about $0.5$\,dB at a BER around $10^{-6}$, and remains within roughly $0.3$\,dB of the Global MMSE upper bound over most of the waterfall region. For $N_R=10$, the gap to this upper bound stays within a small margin (about $0.5$\,dB), while LS remains far behind and LMMSE, although much better than LS, still exhibits a significantly slower BER decay.

The comparison between the two-pilot and one-pilot settings shows that all methods degrade when pilot density is reduced, but the proposed architecture remains relatively robust and preserves most of its advantage over LS, LMMSE, and CNN baselines. A similar trend is observed for the full self-attention Transformer in the cases where it could be evaluated, which confirms that attention-based receivers are well suited to pilot-sparse regimes.

This behavior is also consistent with the physical structure of the problem. Each resource element primarily depends on its surrounding time--frequency neighborhood through the channel coherence structure, while complementary information is available across APs for the same TF position. Accordingly, the shared per-AP encoder progressively exchanges information across TF tokens through successive self-attention layers, and the token-wise cross-attention module performs inter-AP fusion exactly where it is most relevant, i.e., at a fixed TF location across APs. This physically grounded factorization helps explain why the proposed model is both more scalable and slightly more effective than the full self-attention Transformer in the considered setting.

From a spectral-efficiency perspective, reducing the pilot mask from two columns to one increases the fraction of data REs from \(34/36\) to \(35/36\), i.e., a \(2.94\%\) gain.

\begin{table*}[!t]
\centering
\caption{Model size, FLOPs, and CPU latency versus $N_R$ (batch size 1; methods sorted by increasing latency).}
\label{tab:big-flops-lat-params}
\renewcommand{\arraystretch}{1.15}
\setlength{\tabcolsep}{7pt}
%\begin{tabular}{lccccccccc}
\begin{tabular}{llllllllll}
\toprule
& \multicolumn{4}{c}{FLOPs} & \multicolumn{4}{c}{Latency [ms] $\downarrow$} & \\
\cmidrule(lr){2-5}\cmidrule(lr){6-9}
Method
& $N_R{=}1$ & $N_R{=}3$ & $N_R{=}5$ & $N_R{=}10$
& $N_R{=}1$ & $N_R{=}3$ & $N_R{=}5$ & $N_R{=}10$
& Parameters \\
\midrule
LS + Eq. + Demap
  & $2.4{\times}10^{3}$  & $7.2{\times}10^{3}$  & $1.2{\times}10^{4}$  & $2.4{\times}10^{4}$
  & 48  & 165  & 273  & 496
  & N/A \\
CNN\cite{ait_aoudia_end--end_2022}
  & $1.3{\times}10^{10}$ & $4.0{\times}10^{10}$ & $6.6{\times}10^{10}$ & $1.3{\times}10^{11}$
  & 68 & 186 & 302  & 715
  & 8.26\,M \\
LMMSE + Eq. + Demap
  & $1.7{\times}10^{7}$  & $5.0{\times}10^{7}$  & $8.4{\times}10^{7}$  & $1.7{\times}10^{8}$
  & 89  & 276  & 485  & 1012
  & N/A \\
Cross-Attn Transformer (Ours)
  & $3.6{\times}10^{9}$  & $1.1{\times}10^{10}$ & $1.8{\times}10^{10}$ & $3.5{\times}10^{10}$
  & 227  & 651  & 1050 & 2120
  & 0.15\,M \\
Full Self-Attn Transformer\cite{yu_large_2025}
  & $4.3{\times}10^{9}$  & $1.8{\times}10^{10}$ & $3.7{\times}10^{10}$ & $1.1{\times}10^{11}$
  & 272  & 1110 & 2370 & 7700
  & 0.15\,M \\
\bottomrule
\end{tabular}
\end{table*}

% \begin{table*}[!t]
% \centering
% \caption{Model size, GFLOPs and CPU latency versus number of cooperating receivers $N_R$ (batch size 1)}
% \label{tab:big-flops-lat-params}
% \renewcommand{\arraystretch}{1.15}
% \setlength{\tabcolsep}{7pt}
% \begin{tabular}{lccccccccc}
% \toprule
% & \multicolumn{4}{c}{GFLOPs} & \multicolumn{4}{c}{Latency [ms] $\downarrow$} & \\
% \cmidrule(lr){2-5}\cmidrule(lr){6-9}
% Method
% & $N_R{=}1$ & $N_R{=}3$ & $N_R{=}5$ & $N_R{=}10$
% & $N_R{=}1$ & $N_R{=}3$ & $N_R{=}5$ & $N_R{=}10$
% & Parameters \\
% \midrule
% LS + Eq. + Demap
%   & $\sim 10^{-5}$  & $\sim 10^{-5}$  & $\sim 10^{-5}$  & $\sim 10^{-4}$
%   & 48  & 165  & 273  & 496
%   & N/A \\
% CNN\cite{ait_aoudia_end--end_2022}
%   & 13 & 40 & 66 & 130
%   & 68 & 186 & 302  & 715
%   & 8.26\,M \\
% LMMSE + Eq. + Demap
%   & $\sim 10^{-2}$  & $\sim 10^{-2}$  & $\sim 10^{-2}$  & $\sim 10^{-1}$
%   & 89  & 276  & 485  & 1012
%   & N/A \\
% Cross-Attn Transformer (Ours)
%   & 3.6  & 11 & 18 & 35
%   & 227  & 651  & 1050 & 2120
%   & 0.15\,M \\
% Full Self-Attn Transformer\cite{yu_large_2025}
%   & 4.3  & 18 & 37 & 110
%   & 272  & 1110 & 2370 & 7700
%   & 0.15\,M \\
% \bottomrule
% \end{tabular}
% \end{table*}

\subsubsection{Computational complexity and inference time}

The protocol uses 100 inference runs (batch size 1) to obtain stable statistics. For classical methods, this includes channel estimation, equalization, and demapping. For neural models, it corresponds to a forward pass. All measurements are performed on an \textit{AMD Ryzen 5 Pro 7530U}, without GPU acceleration.

Table~\ref{tab:big-flops-lat-params} reports model size, FLOPs, and CPU latency as a function of the number of cooperating receivers $N_R$. Classical baselines (LS and LMMSE) have negligible parameter counts but their FLOPs grow linearly with $N_R$ (one full estimation/equalization/demapping chain per AP), with LMMSE being significantly more expensive than LS.

Among neural models, the self-attention Transformer exhibits the least favorable asymptotic scaling, with a cost that scales as $\mathcal{O}((N_R N_c N_s)^2 d_{\text{model}})$. Per-AP encoders (shared across APs) add a term that scales as $\mathcal{O}(N_R (N_c N_s)^2 d_{\text{model}})$, but this remains dominated by the quadratic self-attention term as $N_R$ grows. The CNN has even higher FLOPs overall, but its convolutions are highly parallelizable on modern hardware, which partly mitigates the wall-clock latency.

In contrast, our method has a much more favorable scaling. The token-wise cross-attention layers scale as $\mathcal{O}(N_R N_c N_s d_{\text{model}})$, i.e., linearly in $N_R$ for a fixed time–frequency grid. In practice, this cost is dominated by the shared per-AP encoder complexity, which scales as $\mathcal{O}(N_R (N_c N_s)^2 d_{\text{model}})$. This results in a lower FLOP count than both the CNN and the full self-attention Transformer, while using only $0.15$M parameters.

The measured inference latency confirms these trends: while the full self-attention Transformer becomes quickly impractical as $N_R$ increases, the proposed model exhibits a much more moderate latency growth. At $N_R=10$, it is over 3$\times$ faster (2120ms vs. 7700ms). This makes it more suitable for future cooperative reception in distributed cell-free deployments. However, it should be noted that these latency figures are meant as indicative comparisons rather than absolute limits: in particular, the classical LS/LMMSE chains are implemented in Python, so their latency is dominated by software overheads rather than pure arithmetic complexity.

\section{Conclusion and perspectives}\label{sec:concl}

We presented a cross-attention Transformer for joint multi-AP uplink decoding, achieving linear complexity in $N_R$ through token-wise fusion across receivers and producing decoder-ready LLRs without explicit CSI estimation. Simulations over 3GPP TR 38.901 UMi channels show consistent gains over LS/LMMSE, CNN, and full self-attention Transformer baselines, while remaining robust to sparse pilots and approaching the Global MMSE upper bound with perfect CSI. The proposed architecture is compact (0.15M parameters, 35 GFLOPs for $N_R=10$) and achieves lower inference latency on commodity CPUs, making it a promising candidate for next-generation distributed Wi-Fi and cell-free architectures. Future work includes multi-user extensions, fronthaul-aware processing under imperfect synchronization, and topology-adaptive fusion.

%\section*{Acknowledgment}
%This work was supported in part by Bpifrance under the France 2030 i‑Démo program (Wi‑FIP project, 2023–2026, Grant I-DEMO-52255).

\bibliographystyle{IEEEtran}
\bibliography{references}

\end{document}